\documentclass[prb,reprint,showpacs,twocolumn,amsmath,amssymb,floatfix]{revtex4}
\usepackage{graphicx}
\usepackage{dcolumn}
\usepackage{bm}
\usepackage{times}
\renewcommand{\text}[1]{\ensuremath{\rm #1}}

\begin{document}
\preprint{}
\title{Universal departure from Johnson-Nyquist relation caused by limited resolution}
\author{Yasuhiro Yamada}
\email[]{yamada@solis.t.u-tokyo.ac.jp}
\author{Masatoshi Imada}
\affiliation{Department of Applied Physics, University of Tokyo, and JST CREST, 7-3-1, Hongo, Bunkyo-ku, Tokyo, 113-8656, JAPAN}

\date{\today}
\begin{abstract}
Exploiting the two-point measurement statistics, we propose a quantum measurement scheme of current with limited resolution of electron counting. Our scheme is equivalent to the full counting statistics in the long-time measurement with the ideal resolution, but is theoretically extended to take into account the resolution limit of actual measurement devices. Applying our scheme to a resonant level model, we show that the limited resolution of current measurement gives rise to a positive excess noise, which leads to a deviation from the Johnson-Nyquist relation. The deviation exhibits universal single-parameter scaling with the scaling variable $Q\equiv S_{\rm{}M}/S_0$, which represents the degree of the insufficiency of the resolution. Here, $S_0$ is the intrinsic noise, and $S_{\rm{}M}$ is the positive quantity that has the same dimension as $S_0$ and is defined solely by the measurement scheme. For the lack of the ideal resolution, the deviation emerges for $Q<1$ as $2\exp[-(2\pi)^2/Q]$ having an essential singularity at $Q=0$, which followed by the square root dependence $\sqrt{Q/4\pi}$ for $Q\gg1$. Our findings offer an explanation for the anomalous enhancement of noise temperature observed in Johnson noise thermometry.
\end{abstract}

\pacs{73.63.-b, 72.70.+m}
\maketitle
\section{Introduction}
In general, an ordinary realistic measurement can also be regarded as an information transfer process between the target system and us via a measurement device, where our available information depends on all of them. A study on the device limitations, therefore, contributes to an understanding of what information is really available in the measurement process. Measurement of a current is one of the most standard techniques to obtain the intrinsic information about the target system in the condensed matter physics. Theoretically, the probability distribution of transferred charge obtained in a current measurement is described by the full counting statistics, that was first proposed by Levitov and Lesovik~\cite{Levitov:1993ma,Levitov:1996ie} and then has been established in the last two decades. Most of theoretical studies, however, focus on the ideal measurement (see Refs~\onlinecite{Nazarov:2003,Esposito:2009zz} and references therein) and only a few of studies deal with the influence of the device limitations~\cite{Naaman:2006,Utsumi:2010,Bednorz:2008}.

When ideal current measurements are conducted, the universal relation is satisfied between the linear conductance and current noise, i.e. the Johnson-Nyquist (J-N) relation~\cite{Johnson:1927tu,Nyquist:1928wx}. The J-N relation is an early significant example of the fluctuation-dissipation theorem~\cite{Callen:1951wg,Kubo:1957wk}, and provides a proportional relation between the variance of a fluctuating current through a conductor, i.e. current noise, and the conductance as
\begin{equation}
S_{0}|_{V=0}=2k_{\rm{}B}TG_{0},
\label{eq:JNrelation}
\end{equation}
where $T$ is the temperature of the conducting electrons, $k_{\rm{}B}$ is the Boltzmann constant, $S_0|_{V=0}$ represents the equilibrium noise, and $G_0\equiv\lim_{V\to0}dI_0/dV$ reads the linear response of the averaged current $I_{0}$ to applied bias voltage $V$, respectively.

In addition to its importance in fundamental physics, the J-N relation also has a practical significance in thermometry~\cite{White:1996wr}. Since the temperature can be determined by measuring only $S_{0}|_{V=0}$ and $G_{0}$, the Johnson noise thermometry has been exploited in rapidly developing noise measurements of nanosystems from which we obtain the useful information about the low-energy excitations in the quantum systems~\cite{Reznikov:1995us,depicciotto:1997dk,Saminadayar:1997tl,Lefloch:2003fp,Sela:2006kq,Zarchin:2008gq,Hashisaka:2008ef,Delattre:2009,Yamauchi:2011cq} and confirm the steady state fluctuation theorem~\cite{Tobiska:2005ht,Saito:2008hs,Nakamura:2010hn}. 

When a sample is placed in a dilute refrigerator, however, the noise temperature determined from the J-N relation, $T_{\rm{}JN}$, is sometimes higher than the temperature of the refrigerator independently measured with a resistance thermometer, $T_{\rm{}ref}$~\cite{Hashisaka:2008ef,Nakamura:2010hn}. The discrepancy has been recognized since early 1970s~\cite{Webb:1973ej}, and attributed to a heat leak to the sample in the refrigerator~\cite{Hashisaka:2008ef,Webb:1973ej}. Since an increasing discrepancy is observed only at very low temperatures above which $T_{\rm{}JN}\simeq{}T_{\rm{}ref}$ is satisfied, it is generally agreed that the measured noise is properly calibrated and $T_{\rm{}JN}$ represents the actual electron temperature~\cite{Nakamura:2010hn}. The seemingly correct interpretation, however, does not include consideration of the possibility of an extrinsic noise enhancing only at such very low temperatures.

In this paper, we theoretically investigate the influence of resolution, (in other words the smallest detectable change in measurement), on the current measurement, which at least qualitatively accounts for the discrepancy. The resolution fundamentally limits the available information in the measurement process, which must affect the observed fluctuation and noise. In fact, the limited resolution gives rise to an enhancement of the extrinsic noise only at very low temperatures as discussed in the Sec. V.

Before going into the detail, we briefly explain our formalism and main results. To understand the resolution effects on the current measurement, we exploit the two-point measurement statistics proposed by Esposito, Harbola, and Mukamel~\cite{Esposito:2009zz}. They calculated the probability distribution of the particle-number change $n\equiv{}N'-N$ taking place in a part of the system in a measurement time ${\cal{}T}$. $N$ and $N'$ read the particle numbers of the part at $t=0$ and $t={\cal{}T}$, respectively, which are given by the projective measurement in the basis of the particle-number operator, $\hat{N}_{\rm{}part}$. Note that the equation of continuity connects $n$ with the net current flowing into the part. $n$ can be any integer, which means that the electrons in current are ideally distinguished, one by one. We extend their scheme of current measurement to take into account a limited resolution $\Delta$. In other words, we study a coarse-graining of the available information on current. $\Delta$ is introduced in the particle-number measurements at $t=0$ and $\cal{}T$, which are described by projection operators parameterized by an integer $k$, $\{\hat{P}_{k}^{\rm part}(\Delta)\}$, where
\begin{equation}
\hat{P}_{k}^{\rm part}(\Delta)\equiv\int_{\chi_{k}-\frac{\Delta}{2}}^{\chi_{k}+\frac{\Delta}{2}}dx\delta(x-\hat{N}_{\rm{}part}).
\end{equation}
Here, $\chi_{k}\equiv\chi_{0}+k\Delta$ is the outcome of the measurement where $\chi_0$ is the zero-point deviation. In our scheme, $n\equiv{}\chi_{k'}-\chi_{k}=(k'-k)\Delta$ is the available outcome and can be any multiple of $\Delta$, which means that $\Delta$-particles are required for the detection of the change in $n$ at least.

Our scheme is described by a positive operator-valued measure~\cite{Davies:1970ux,Kraus:1971wd} (POVM) measurement characterized by two measurement parameters, ${\cal{}T}$ and $\Delta$. It is noteworthy that the scheme is reduced to that of Esposito {\it et al}.~\cite{Esposito:2009zz} and the full counting statistics proposed by Levitov and Lesovik~\cite{Levitov:1993ma,Levitov:1996ie} in the case of $\Delta=1$ with a long ${\cal{}T}$ in comparison with the characteristic time scale of the transport in the target system. 

\begin{figure}[tb]
\begin{center}
\includegraphics[width=40mm]{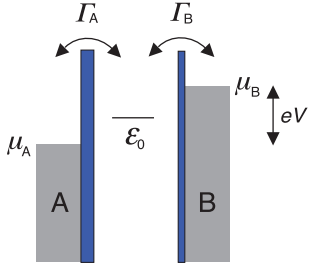}
\end{center}
\caption{(color online). Schematic illustration of resonant level model. The $\varepsilon_{0}$-level is coupled to two reservoirs A and B between which the bias voltage $V$ is applied. $\mathit{\Gamma}_{\rm A(B)}$ reads the characteristic frequency of the electron transfer between the level and the reservoir A(B). $\mu_{\rm A(B)}$ represents the chemical potential of the reservoir A(B). We take $\mu_{\rm{}A}=0$ and $\mu_{\rm{}B}=eV$. We introduce $\mathit{\Gamma}^{-1}\equiv[(\mathit{\Gamma}_{\rm{}A}+\mathit{\Gamma}_{\rm{}B})/2]^{-1}$ and $r\equiv\mathit{\Gamma}_{\rm{}A}\mathit{\Gamma}_{\rm{}B}/\mathit{\Gamma}^2$ as the characteristic time scale and the degree of asymmetry of the couplings, respectively.}
\label{fig1}
\end{figure}

Since the available information depends on the measurement device, it is important to explain what is our intended device. As a model for actual galvanometers, Levitov and Lesovik introduced a precessing 1/2 spin, which measures a current indirectly via the induced magnetic field~\cite{Levitov:1996ie}: The precession angle is proportional to the net charge transferred near by the spin for a measurement time, $\cal{}T$. Our scheme is, therefore, expected to take into account the essence of a conventional current-measuring device including the function of a galvanometer, which requires $\Delta$-electrons at least during a time ${\cal T}$ to work. Note that in our scheme, most of the electrons can move without disturbance by projection during the measurement because $\cal{}T$ is usually much longer than the microscopic time scale of electrons. In contrast to the conventional current measurement, a newly developing charge-sensing device, a quantum-point-contact detector, works in a different way and gives us a real-time detection of a charge state by projecting the system to the charge diagonal state~\cite{Naaman:2006,Utsumi:2010,Fujisawa:2006jf,Gustavsson:2006jm,Gustavsson:2007,Kung:2012ct}. Namely, our scheme describes the conventional current measurement device but the newly developing one.

\begin{figure}[tb]
\begin{center}
\includegraphics[width=86mm]{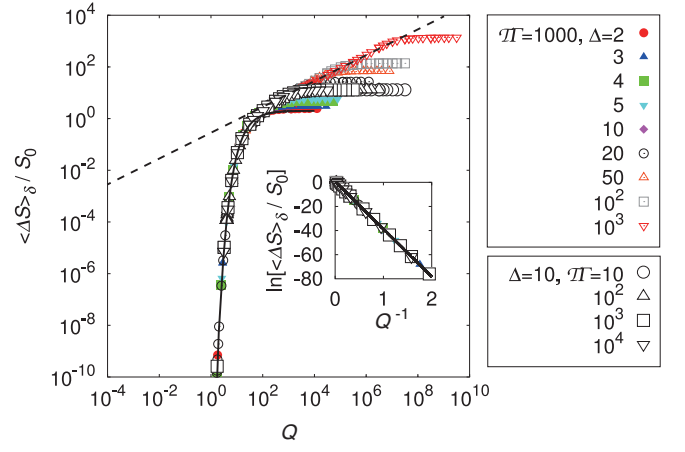}
\end{center}
\caption{
(color online). Ratio of excess and intrinsic noises $\langle\mathit{\Delta}S\rangle_{\delta}/S_{0}$ in the thermal equilibrium state ($V=0$) as a function of $Q\equiv S_{\rm{}M}/S_{0}$ for several choices 
of (${\cal{}T},\Delta$), where $S_{\rm{}M}\equiv(e\Delta)^2/{\cal{}T}$. The other parameters are fixed at $\varepsilon_0=0$ and $r=1$. The black solid line indicates the universal exponential $A\exp[-\gamma/Q]$, with $A=2$ and $\gamma=(2\pi)^2$ estimated from Eq.~\eqref{eq:estimation_exponential}. The dashed line represents the square root dependence $B\sqrt{Q}$, where $B=1/\sqrt{4\pi}$ determined from Eq.~\eqref{eq:estimation_squareroot}. The inset shows the linear dependence of the logarithm of the ratio on $Q^{-1}$.
}
\label{fig2}
\end{figure}

Applying the extended two-point measurement scheme to the current through a resonant level depicted in Fig.~\ref{fig1}, we show that the limited resolution gives rise to the departure of the measured noise $S$ from the intrinsic one $S_{0}$ while the measured current $I$ is unchanged at $I_{0}$. The excess noise, $\langle\mathit{\Delta}S\rangle_{\delta}=S-S_0$, is positive and shows an anomalous temperature dependence, which can make the usual empirical method of noise calibration~\cite{DiCarlo:2006} unjustified~\cite{Hashisaka:2008ef,Nakamura:2010hn,Webb:1973ej}. Note that $\langle\mathit{\Delta}S\rangle_{\delta}$ is explicitly evaluated by using Eq.~\eqref{eq:averagednoise}. Hence, the J-N relation can be violated between the measured noise $S$ and measured conductance $G\equiv\lim_{V\to0}dI/dV$ in the practical cases, which causes a discrepancy between $T_{\rm JN}$ and $T_{\rm ref}$ at low temperatures. The deviation from the J-N relation between $S$ and $G$ caused by the limited resolution is represented by,
\begin{equation}
\frac{S|_{V=0}}{2k_{\rm{}B}TG}-1=\frac{\langle\mathit{\Delta}S\rangle_{\delta}}{S_{0}}\Big|_{V=0}\ge{}0.\label{eq:deviationfromJN}
\end{equation}
It is remarkable that the ratio of noises obeys a scaling law with the scaling variable $Q\equiv S_{\rm{}M}/S_{0}$ as seen in Fig.~\ref{fig2}, where $S_{\rm{}M}\equiv(e\Delta)^2/{\cal{}T}$ is the characteristic noise determined solely from the measurement scheme. The scaling function exhibits the universal exponential dependence for $Q<1$ having the essential singularity at $Q=0$ with increasing from zero to unity, and shows a crossover to an algebraic increase or a constant at $Q>1$. From the scaling law, we find that $S_0$ is not detectable in noise experiments when $S_0$ is much smaller than $S_\textrm{M}$, $Q\gg1$. Since $\Delta_{0}\equiv\sqrt{S_0\mathcal{T}}/e$ is the standard deviation of the transferred particle number counted with the ideal resolution, it means the average number of particle involved in the measurement for $V=0$. The enhanced deviation for large $Q=(\Delta/\Delta_{0})^2$ is, therefore, consistent with our intuition that the resolution error of noise should be more profound in the case that only a few particles are involved. Although the above discussion of scaling is based on the specific model, essentially the same scaling relation is expected be satisfied for an arbitrary mesoscopic conductor coupled to normal reservoirs, as will be discussed in Sec. IV. The experimental anomalous enhancement of noise at low temperatures can be understood by the scaling behavior: The excess noise being irrelevant at high temperatures becomes profound at low temperatures because $Q$ increases with decreasing the temperature. Note that there are other known noise sources that make the violation of the J-N relation, e.g. the background noise. The noises coming from the sources, however, can be calibrated by using the empirical method because of their trivial temperature dependences accounted for by circuit theory~\cite{DiCarlo:2006} and do not give an explanation for the observed discrepancy between $T_{\rm{}JN}$ and $T_{\rm{}ref}$.

The plan of the paper is the following. In Sec. II we formulate the resolution of the current measurement exploiting the two-point measurement, and obtain a formula which describes the characteristic function of the distribution of the transferred particle number counted with limited resolution. In Sec. III, we apply the formula to the resonant level model and calculate the measured current and measured noise analytically. Section IV gives the numerical calculations of the intrinsic and excess noises in the thermal equilibrium state and the linear response of the current. Section V is devoted to the comparison between theory and experiment. It is clarified that our results are consistent with the experiments and may account for the difference between $T_{\rm{}JN}$ and $T_{\rm{}ref}$. A summary and conclusions of our work are given in Sec. VI.

\section{Formalism of Current Measurement with Limited Resolution}
In this section, we formulate the two-point measurement statistics under limited resolutions of steady state current through a reservoir (lead) in a multi-terminal mesoscopic system that consists of a conductor connected to multiple reservoirs. The system is described by the following general Hamiltonian,
\begin{align}
\hat{\cal{}H}(t)=\hat{H}_{0}+\hat{V}(t),
\label{eq:Hamiltonian}
\end{align}
where
\begin{align}
\hat{H}_{0}&=\hat{H}_{\rm{}con}+\sum_{\rm X=A,B,\cdots}\hat{H}_{\rm{}X},\\
\hat{V}(t)&=\sum_{\rm X=A,B,\cdots}\hat{V}_{\rm{}X}\theta(t).
\end{align}
Here $\hat{H}_{\rm{}con}$ and $\hat{H}_{\rm{}X}$ read the Hamiltonians of the conductor and the reservoir X, respectively, $\hat{V}_{\rm{}X}$ is the hopping matrix between the reservoir X and the conductor, and $\theta(t)$ is the step function.

The current is observed as the net change of particle number in the reservoir A from $t=0$ to $t={\cal T}$. Before the current measurement, it is assumed that the conductor is disconnected for $t\le 0$ from all of the reservoirs, which are in the isolated thermal equilibrium states with the different chemical potentials. Then, the density matrix at $t=0$ is given by 
\begin{align}
\hat{\rho}(0)&=\hat{\rho}_{{\rm con}}^{0}\otimes\frac{\exp[-\beta(\hat{H}_{{\rm A}}-\mu_{{\rm A}}\hat{N}_{{\rm A}})]}{\textrm{Tr}\Big[\exp[-\beta(\hat{H}_{{\rm A}}-\mu_{{\rm A}}\hat{N}_{{\rm A}})]\Big]}\notag\\
&\quad\otimes \frac{\exp[-\beta(\hat{H}_{{\rm B}}-\mu_{{\rm B}}\hat{N}_{{\rm B}})]}{\textrm{Tr}\Big[\exp[-\beta(\hat{H}_{{\rm B}}-\mu_{{\rm B}}\hat{N}_{{\rm B}})]\Big]}\otimes \cdots,
\end{align}
where $\hat{N}_{{\rm X}}$ is the total number operator of the reservoir $X$ that commutes with $\hat{H}_{\rm X}$, $\beta\equiv 1/k_{\rm{}B}T$ is the inverse temperature of the system, $\hat{\rho}_{{\rm con}}^{0}$ is the initial density matrix of the conductor, and $\mu_{\rm X}$ represents the chemical potential of the reservoir X. Since the reservoir A is isolated for $t\le 0$, the particle number of the reservoir A takes a constant, $N_{{\rm A}}^0$, which is the initial particle number of the reservoir A at $t=0$: $\hat{\rho}(0)\hat{N}_{{\rm A}}=N_{{\rm A}}^{0}\hat{\rho}(0)$. It is noteworthy that any number of channels of the reservoir and any interaction of the conductor, e.g. Coulomb interaction, can be dealt with in this model.

Our measurement scheme is a simple extension of that proposed by Esposito, Harbola, and Mukamel~\cite{Esposito:2009zz}. Note that in Ref. \onlinecite{Esposito:2009zz}, the full counting statistics is reformulated with using the superoperators in Liouville space, that is convenient to the simple description of the current measurement scheme. We here, however, use the ordinary operators in Hilbert space for the convenience of the general readers.

The indirect measurement of current flowing into the reservoir A via the induced magnetic field can be described by the measurement of the number of electrons flowing into reservoir A during a measurement time, $\cal{}T$. Esposito, Harbola, and Mukamel calculated the probability that the slight change in the particle number in the reservoir A during a measurement time ${\cal{}T}$ is equal to $k$ with the following two-point measurement,
\begin{equation}
{\cal P}^{\rm{}EHM}(k;{\cal T})= \sum_{l}\textrm{Tr}[\hat{P}_{l+k}\hat{U}({\cal T},0)\hat{P}_{l}\hat{\rho}(0)\hat{P}_{l}\hat{U}^\dagger({\cal T},0)\hat{P}_{l+k}],
\label{eq:P_EHM}
\end{equation}
where $\hat{P}_{k}\equiv|k\rangle\langle{}k|$ is the projective operator of the particle number operator of reservoir A, $\hat{N}_{A}=\sum_{k}k|k\rangle\langle{}k|$, where $k$ is the eigenvalue, and $\hat{U}(t,t')\equiv\breve{T}\exp\big[-\frac{i}{\hbar} \int_{t'}^{t}\hat{\cal H}(t_1) dt_1\big]$ reads the time-evolution operator. They showed that the cumulant generating function of ${\cal P}^{\rm{}EHM}(k;{\cal T})$ is equal to the one obtained in the full counting statistics in the case of ${\cal T}\mathit{\Gamma}\gg1$. From the viewpoint of quantum measurement theory, the measurement can be described by the POVM formalism,
\begin{equation}
{\cal{}P}^{\rm{}EHM}(k;{\cal{}T})={\rm{}Tr}[\hat{D}_{k}^{\rm{}EHM}({\cal{}T})\hat{\rho}(0)],
\end{equation}
where the operators $\hat{D}_{k}^{\rm{}EHM}({\cal{}T})$ are the POVM elements defined by
$\hat{D}_{k}^{\rm{}EHM}({\cal{}T})\equiv \sum_{l}\hat{M}_{k,l}^{\rm{}EHM\dagger}({\cal T})\hat{M}_{k,l}^{\rm{}EHM}({\cal T})$ where
\begin{equation}
\hat{M}_{k,l}^{\rm{}EHM}({\cal T})\equiv \hat{P}_{l+k}\hat{U}({\cal T},0)\hat{P}_{l}.
\end{equation}

In their calculation, the outcome of ${\cal P}^{\rm{}EHM}(k;{\cal T})$, $k$, can be any integers, which implies that the measurement device has the function to detect the change of even just one electron during $\cal{}T$. That is, however, not realistic. The ultimately high resolution is attributed to the part of the projective measurement, $\hat{P}_{k}$.

We implement the limitation of the resolution by introducing smallest detectable number of electrons $\Delta$ and replace $\hat{P}_{k}$ with a projection operator $\hat{P}_{k}(\Delta)$ defined by
\begin{equation}
\hat{P}_{k}(\Delta)\equiv \int_{\chi_{k}-\frac{\Delta}{2}}^{\chi_{k}+\frac{\Delta}{2}}dx \delta(x -\hat{N}_{A}).
\end{equation}
Here, $\chi_k\equiv\chi_{0}+k\Delta-\eta$. $\chi_0$ and $\eta$ read the zero point deviation of the particle-number measurement and the positive infinitesimal, respectively. $\hat{P}_{k}(\Delta)$ satisfies $\hat{P}_{k}(\Delta)\hat{P}_{l}(\Delta)=\delta_{k,l}\hat{P}_{k}(\Delta)$ and projects a state onto the subspace spanned by the eigenvectors belonging to the eigenvalues of $\hat{N}_{\rm{}A}$ which satisfy $\chi_{k}-\frac{\Delta}{2}\le N_{A} <\chi_{k}+\frac{\Delta}{2}$. $\Delta$, therefore, represents the resolution of the particle-number measurement of the reservoir A and becomes a scale unit in the outcome.

With using the projection operators, the probability that the particle number change of the reservoir A during ${\cal T}$ is equal to $k\Delta$, ${\cal P}(k;{\cal T},\Delta)$, is obtained from 
\begin{equation}
{\cal P}(k;{\cal T},\Delta)= \textrm{Tr}[\hat{D}_{k}({\cal T},\Delta)\hat{\rho}(0)],
\label{eq:P}
\end{equation}
where $\hat{D}_{k}({\cal T},\Delta)\equiv \sum_{l} \hat{M}_{k,l}^{\dagger}({\cal T},\Delta)\hat{M}_{k,l}({\cal T},\Delta)$ are POVM~\cite{Davies:1970ux,Kraus:1971wd} elements. The operators $\hat{M}_{k,l}({\cal T},\Delta)$ are defined by the following equation;
\begin{equation}
\hat{M}_{k,l}({\cal T},\Delta)\equiv \hat{P}_{l+k}(\Delta)\hat{U}({\cal T},0)\hat{P}_{l}(\Delta).
\end{equation}
Note that although, in this paper, we consider the particle flow with the two-point measurement statistics with a limited resolution, our definition of resolution is easy to be extended and can be applied to the measurement of other physical quantities such as heat current. In that case, the resolution could be more significant because there is no apriori unit of the measurement.

For the calculation of the average and the variance of the current, it is useful to consider the characteristic function of the probability defined by ${\cal M}(\lambda;{\cal T},\Delta)\equiv \sum_{k}\exp[i \lambda k]{\cal P}(k;{\cal T},\Delta)$. With some calculations, the characteristic function is written as
\begin{align}
&{\cal M}(\lambda;{\cal T},\Delta)\notag\\
&=\sum_{m=-\infty}^{\infty}\mathrm{sinc}(\frac{\lambda+2\pi m}{2})\exp[i2\pi m \frac{\delta}{\Delta}]{\cal M}_{0}(\frac{\lambda+2\pi m}{\Delta}, {\cal T}),
\label{eq:momentGF}
\end{align}
where
\begin{equation}
{\cal M}_{0}(\lambda; {\cal T})\equiv \textrm{Tr}[\hat{U}^\dagger({\cal T},0;-\frac{\lambda}{2})\hat{U}({\cal T},0;\frac{\lambda}{2}) \hat{\rho}(0)],
\end{equation}
\begin{equation}
\delta \equiv N_{{\rm A}}^0-\chi_{0} \bmod \Delta \quad (0 \le \delta < \Delta).
\end{equation}
$\hat{U}(t,t';\lambda)\equiv \breve{T}\exp[-i/\hbar \int_{t'}^{t}\hat{\cal H}(t_1;\lambda) dt_1]$ is the modified time evolution operator with the counting field $\lambda$ where $\hat{\cal H}(t;\lambda)\equiv \exp[i\lambda\hat{N}_{{\rm A}}]\hat{\cal H}(t)\exp[-i\lambda\hat{N}_{{\rm A}}]$, and $\mathrm{sinc}(x)\equiv \sin(x)/x$. Note that in the above calculation, we ignore a constant factor of ${\cal M}(\lambda;{\cal T},\Delta)$ which does not affect our final results.

In Eq.~\eqref{eq:momentGF}, all the detailed information of the target system is included in ${\cal M}_{0}(\lambda; {\cal T})$ that is the characteristic function of the distribution of the transferred particle number in the ideal resolution case. Equation \eqref{eq:momentGF} represents, therefore, the general formula of the characteristic function of the transferred particle number counted with the limited resolution.

\section{Application to Resonant Level Model and Random Averaging}
To proceed the concrete calculation, we apply the above formal result to the resonant level connected to two noninteracting reservoirs (see Fig.~\ref{fig1}). The Hamiltonian of the resonant level model which consists of a resonant level $\varepsilon_{0}$ coupled to two reservoirs A and B is represented by Eq. \eqref{eq:Hamiltonian} with replacing the terms with $\hat{H}_{0}=\hat{H}_{\rm{}A}+\hat{H}_{\rm{}B}+\hat{H}_{\rm{}sys}$, $\hat{V}(t)=\hat{V}_{\rm{}A}\theta(t)+\hat{V}_{\rm{}B}\theta(t)$, $\hat{H}_{\rm{}sys}=\varepsilon_{0}\hat{d}^\dagger\hat{d}$, $\hat{H}_{\rm{}X}=\sum_{x{}\in{\rm{}X}}\varepsilon_{x}^{\rm{}X}\hat{c}_{x}^\dagger\hat{c}_{x}$, and $\hat{V}_{X}=\sum_{x{}\in{\rm{}X}}(t_{\rm{}X}\hat{d}^\dagger\hat{c}_{x}+{\rm{}H.c.})$ for ${\rm{}X}={\rm{}A},{\rm{}B}$. Here, $\hat{d}^{\dagger}$ creates a spinless electron with charge $e$ at the resonant level $\varepsilon_{0}$, while $\hat{c}_{x\in\text{X}}^{\dagger}$ denotes the creation operator of a spinless electron at a wave number $x$ in the reservoir X=A or B, with a constant density of states $\rho_{\text{X}}$. The resonant level is coupled to the reservoir X with a hybridization $t_{\text{X}}$, where the characteristic transport frequency $\mathit{\Gamma}_X$ is given by $\mathit{\Gamma}_{\rm{}X}=2\pi|t_{\rm{}X}|^2\rho_{\rm{}X}/\hbar$. The chemical potentials of reservoirs have the different values, $\mu_{B}=eV$ and $\mu_{A}=0$, because of the applied bias voltage $V$ between the reservoirs. We note that though the reservoir A is used for the two-point measurement, the choice of the reservoir does not influence our results in this two-terminal case. 

To obtain the stationary current distribution, ${\cal{}T}$ is assumed to be much longer than the characteristic time scale of the electrons determined by $\mathit{\Gamma}^{-1}\equiv[(\mathit{\Gamma}_{\rm{}A}+\mathit{\Gamma}_{\rm{}B})/2]^{-1}$ but finite. This model can be considered as a simple model of a quantum dot coupled to two reservoirs, which is one of the typical nanosystems where the noise measurements have been conducted at very low temperatures in the experimental studies~\cite{Zarchin:2008gq,Delattre:2009,Yamauchi:2011cq}. In addition, our model in the equilibrium state with $k_{\rm{}B}T/\hbar\mathit{\Gamma}\ll1$ also describes a single-channel quantum point contact (QPC)~\cite{Hashisaka:2008ef} where the transmission probability is given by $r/[(\varepsilon_0/\hbar\mathit{\Gamma})^2+1]$. Here, $r\equiv\mathit{\Gamma}_{\rm{}A}\mathit{\Gamma}_{\rm{}B}/\mathit{\Gamma}^2$ represents the coupling asymmetry.

Being described by the forward and backward time-evolutions obeying the different modified Hamiltonians, $\hat{\cal H}(t;\pm\lambda/2)$, ${\cal M}_{0}(\lambda;{\cal{}T})$ in Eq.~\eqref{eq:momentGF} is adequately evaluated with using the Keldysh Green's function method~\cite{Kindermann:2003th,Kamenev:2005vu}. ${\cal T}\mathit{\Gamma}\gg1$ is necessary for measuring the stationary current statistics. The leading time order for logarithm of ${\cal M}_{0}(\lambda; {\cal T})$ is evaluated as 
\begin{align}
\ln{\cal M}_{0}(\lambda; {\cal T})={\cal T}\mathit{\Gamma}{\cal C}_{0}(\lambda)+o(\mathcal{T})
\label{eq:longtimeapproximation}
\end{align}
where 
\begin{align}
{\cal C}_{0}(\lambda)&\equiv\int_{-\infty}^{\infty}\frac{dx}{2\pi}\ln\Big[1+T(x)\big[(\exp[i\lambda]-1)[1-f_{\rm A}(x)]f_{\rm B}(x)\notag\\
&\quad+(\exp[-i\lambda]-1)f_{\rm A}(x)[1-f_{\rm B}(x)]\big]\Big]
\end{align}
is the cumulant generating function of current obtained with the Levitov-Lesovik formula~\cite{Levitov:1993ma,Esposito:2009zz}. It is noteworthy that the steady state current statistics is determined solely from the leading order. Hence we omit the sub-leading order terms that describe the approach from the disconnected state at $t=0$ to the connected state where the steady state current flows.
Here $T(x)\equiv r/[(x-\varepsilon_{0}/\hbar\mathit{\Gamma})^2+1]$ reads the transmission probability of the system, and $f_{\rm X}(x)\equiv [\exp[\beta\hbar\mathit{\Gamma}(x-\mu_{\rm X}/\hbar\mathit{\Gamma})]+1]^{-1}$ is the Fermi-Dirac distribution function for the reservoir $\rm X$. We then obtain the following asymptotic form of the characteristic function:
\begin{align}
{\cal M}(\lambda;{\cal T},\Delta)&= \sum_{m=-\infty}^{\infty}\mathrm{sinc}(\frac{\lambda+2\pi m}{2})\exp[i2\pi m \frac{\delta}{\Delta}]\notag\\
&\quad\times\exp\Big[{\cal T}\mathit{\Gamma}{\cal C}_{0}(\frac{\lambda+2\pi m}{\Delta})\Big].
\label{eq:characteristic}
\end{align}

In Eq.~(\ref{eq:characteristic}), ${\cal M}(\lambda;{\cal T},\Delta)$ depends on $\delta$, which means that we can in principle distinguish each specific initial state with the ideal resolution. The distinction, however, blurs in actual experiments. To take into account the actual resolution limit for initial preparation, we take a simple average over $\delta$ for $\ln{\cal M}(\lambda;{\cal T},\Delta)$ as
\begin{equation}
\langle {\cdots}\rangle_{\delta}\equiv \int_{0}^{\Delta}\frac{d\delta}{\Delta} \cdots.
\end{equation}
We assume that the $\delta$-averaging appropriately simulates actual current measurements because it is hardly possible that the current is repeatedly measured under an identical condition with a fixed $\delta$. In other words, the $\delta$-averaging of the logarithm of ${\cal M}(\lambda;{\cal T},\Delta)$ is an analogy of the random average in quenched random systems.

Accordingly, the cumulant generating function of the particle current in the long time measurement is given by
\begin{equation}
{\cal{}C}_{I}(\lambda;{\cal{}T},\Delta)=\frac{\partial\langle\ln{\cal{}M}(\lambda;{\cal{}T},\Delta)\rangle_\delta}{\partial\cal{}T},
\label{eq:CGF}
\end{equation}
In the case of $\Delta=1$, the cumulant generating function in Eq. (\ref{eq:CGF}) is identical to that obtained in the previous study, ${\cal{}C}_{I}(\lambda;{\cal{}T},1)={\cal{}C}_{0}(\lambda)$~\cite{Esposito:2009zz}.

Here, we focus on the averaged current $I$ and the noise $S$ measured by the above measurement scheme. By differentiating the cumulant generating function ${\cal{}C}_{I}(\lambda;{\cal{}T},\Delta)$ in terms of $\lambda$, we evaluate $I$ and $S$ as
\begin{equation}
I=e\Delta\frac{\partial {\cal C}_{I}(\lambda,{\cal T},\Delta)}{\partial (i\lambda)}\Big|_{\lambda=0}= I_{0}+\langle \mathit{\Delta} I \rangle_\delta,
\label{eq:observedcurrent}
\end{equation}
\begin{equation}
S=e^2\Delta^2\frac{\partial^2 {\cal C}_{I}(\lambda,{\cal T},\Delta)}{\partial (i\lambda)^2}\Big|_{\lambda=0}= S_{0}+\langle \mathit{\Delta} S \rangle_\delta,
\label{eq:observednoise}
\end{equation}
where $I_{0}\equiv e\mathit{\Gamma}\partial {\cal C}_0(\lambda)/\partial(i\lambda)|_{\lambda=0}$ and $S_{0}\equiv e^2\mathit{\Gamma}\partial^2 {\cal C}_0(\lambda)/\partial(i\lambda)^2|_{\lambda=0}$ are the intrinsic current and the intrinsic noise obtained in the ideal measurement case of $\Delta=1$, respectively. $I_{0}$ and $S_{0}$ are determined only by the intrinsic parameters of the system and which satisfy the J-N relation. The excess terms, attributed to the limited resolution measurement, can be evaluated as
\begin{equation}
\langle\mathit{\Delta}I\rangle_{\delta}=0,
\label{eq:I0}
\end{equation}
and
\begin{equation}
\langle\mathit{\Delta}S\rangle_{\delta}=-\frac{e^2\mathit{\Gamma}\Delta^2}{2 \pi^2}\sum_{m\ge 1}\frac{\exp\big[{\cal T}\mathit{\Gamma}{\cal C}_{0}^{{\rm sym}}(\frac{2\pi m}{\Delta})\big]{\cal C}_{0}^{{\rm sym}}(\frac{2\pi m}{\Delta})}{m^2}.
\label{eq:averagednoise}
\end{equation}

Here we define ${\cal C}_{0}^{{\rm sym}}(\lambda)\equiv{\cal C}_{0}(\lambda)+{\cal C}_{0}(-\lambda)$. Equation (\ref{eq:I0}) agrees with the naive intuition that the intrinsic current is correctly obtained for the repeated measurement. Note that $\langle \mathit{\Delta} S \rangle_\delta$ depends on the measurement parameters, ${\cal T}$ and $\Delta$, as well as the parameters of the system. From this result, it is found that the limited resolution does not affect the average of the current, which means that our measurement scheme is unbiased. In addition, it is remarkable that the excess noise is always non-negative,
\begin{equation}
\langle \mathit{\Delta} S \rangle_\delta\ge 0,
\label{eq:DeltaSPositive}
\end{equation}
because ${\cal C}_{0}^{{\rm sym}}(\lambda)\le{}0$. These results are general for any $V$.

In the case of $\Delta=1$, since ${\cal C}_{0}^{\rm sym}(2\pi m)=0 $, the excess noise obviously disappears in accordance with our expectation that the measured noise and measured current satisfy the J-N relation in the ideal case.
On the other hand, for large $\Delta$, Eq. (\ref{eq:averagednoise}) is evaluated as 
\begin{equation}
\langle\mathit{\Delta} S\rangle_{\delta}\approx\frac{e^2\mathit{\Gamma}\Delta}{2 \pi^2}\int_{0}^{\infty}s(x,{\cal T})dx
\label{eq:lineardelta}
\end{equation}
where
\begin{equation}
s(x,{\cal T})\equiv -\frac{\exp\big[{\cal T}\mathit{\Gamma}{\cal C}_{0}^{{\rm sym}}(2\pi x )\big]{\cal C}_{0}^{{\rm sym}}(2\pi x)}{x^2}.
\end{equation}
Since $s(x,{\cal T})$ is independent of $\Delta$, the excess noise scales linearly with large $\Delta$.

Here we explain the origin of the excess terms, $\langle\mathit{\Delta}I\rangle_{\delta}$ and $\langle\mathit{\Delta}S\rangle_{\delta}$. These terms can be regarded as the resolution error because it vanishes at $\Delta=1$ and depend on the measurement parameters and $\delta$. $\delta\equiv{}N_A^0-\chi_0\bmod{}\Delta$ represents the degree of freedom for the initial particle number of the reservoir A hidden in the limited resolution. The vanishing excess currents and the non-negative excess noise by the $\delta$-averaging, therefore, mean that the lack of our knowledge of the initial conditions beyond the resolution makes the cancellation of the excess current, namely no error on average, and enhances the measured noise.

Note that essentially the same Equations~(\ref{eq:observedcurrent}-\ref{eq:averagednoise}) can be obtained not only for the present resonant level model but also for general mesoscopic systems which obey the Hamiltonian \eqref{eq:Hamiltonian} when we assume that the leading time order of $\ln\mathcal{M}_{0}(\lambda;\mathcal{T})$ is proportional to ${\cal T}$. This assumption is physically sound when the steady-state exists in the mesoscopic systems because the transferred particle number during the measurement time $\mathcal{T}$ should be proportional to $\mathcal{T}$ for the long time measurement. The coefficient of the term proportional to $\mathcal{T}$ is given by the cumulant generating function of the steady-state current measured with the ideal resolution, as shown in Eq.~\eqref{eq:longtimeapproximation}. The assumption is valid even for the quantum dot systems which include the Coulomb interaction effects~\cite{Bagrets:2003,Utsumi:2006} and the electron-phonon couplings.~\cite{Avriller:2009}

\section{Numerical Results in the thermal equilibrium}
\begin{figure}[tb]
\begin{center}
\includegraphics[width=86mm]{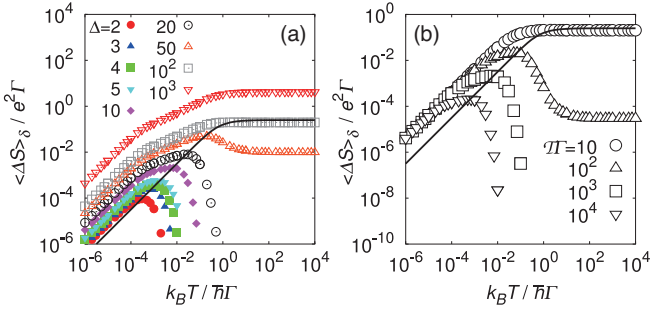}
\end{center}
\caption{
(color online). Equilibrium excess noise $\langle\mathit{\Delta}S\rangle_{\delta}$ at $V=0$ as a function of temperature for $\varepsilon_0=0$ and $r=1$. The measurement parameters are fixed at ${\cal{}T}\mathit{\Gamma}=1000$ in (a) and $\Delta=10$ in (b). The solid line indicates the equilibrium intrinsic noise, $S_{0}/e^2\mathit{\Gamma}$.}
\label{fig3}
\end{figure}

Hereafter, we focus on the equilibrium noises and the linear conductance, $G\equiv\lim_{V\to0}dI/dV=G_0$, in the resonant level model to discuss the resolution effects on the J-N relation. For simplicity, $S$, $S_0$, and $\langle\mathit{\Delta}S\rangle_{\delta}$ are always assumed to carry the measured, intrinsic, and excess noises at $V=0$, respectively. Figure \ref{fig3}(a) shows $\langle\mathit{\Delta}S\rangle_{\delta}$ as a function of temperature $T$ for several choices of $\Delta$. Let us first consider $\Delta<50$. With decreasing $T$, the excess noise $\langle\mathit{\Delta}S\rangle_{\delta}$ increases and shows a peak at a temperature $k_{\rm{}B}T<\hbar\mathit{\Gamma}$ where $S_{0}$ decreases proportionally to $T$. This means that the excess noise may appear only at sufficiently low temperatures. The appearance leads to a difficulty in measuring the intrinsic noise in experiments. With an increase in $\Delta$, $\langle\mathit{\Delta}S\rangle_{\delta}$ is enhanced, and becomes pronounced even at high temperatures $k_{\rm{}B}T\gg\hbar\mathit{\Gamma}$. The measurement time ${\cal T}$ also affects $\langle\mathit{\Delta}S\rangle_{\delta}$ as seen in Fig.~\ref{fig3}(b) where $\langle\mathit{\Delta}S\rangle_{\delta}$ is suppressed with an increase in ${\cal{}T}$. The larger $\cal{}T$ is, therefore, the smaller intrinsic noise we can access in the experiments.

To investigate the resolution effects on the J-N relation in more detail, we calculate the ratio of excess and intrinsic noises which characterizes the deviation from the J-N relation between $S$ and $G$, namely $\langle\mathit{\Delta}S\rangle_{\delta}/S_{0}$ in Eq.~\eqref{eq:deviationfromJN}, as has been already shown in Fig~\ref{fig2}. Figure~\ref{fig2} illustrates the ratio $\langle\mathit{\Delta}S{}\rangle_{\delta}/S_{0}$ as a function of a single non-dimensional positive parameter $Q=S_{\rm{}M}/S_{0}$ for several choices of (${\cal{}T}$, $\Delta$). A scaling behavior is found in the deviation from the J-N relation. All the curves collapse into a single one for $Q\ll10^2$, that is described by the exponential dependence 
\begin{align}
\langle\mathit{\Delta}S\rangle_{\delta}/S_{0}=A\exp[-\gamma/Q].
\label{eq:exponentialscaling}
\end{align}
Above $Q\approx{}10^2$, there exists another single-parameter scaling described by
\begin{align}
\langle\mathit{\Delta}S\rangle_{\delta}/S_{0}=B\sqrt{Q}.
\label{eq:squarerootscaling}
\end{align}
Here, $A$, $\gamma$ and $B$ are estimated as $A=2$, $\gamma=(2\pi)^2$ and $B=1/\sqrt{4\pi}$ from the analytical discussion in the later part of this section below Eq.~\eqref{eq:generalexcessnoise}. Then all the curves saturate at a sufficiently high $Q$, the saturated value of which is not universal but roughly scaled by $\Delta$. The saturation occurs roughly at the crossing of $\Delta$ and $B\sqrt{Q}$ as $Q\approx{}4\pi\Delta^2$. The deviation, therefore, becomes serious at low temperatures and for low conductance which satisfies $Q=S_{\rm{}M}/2k_{\rm{}B}TG>(2\pi)^2/\ln 2\simeq 56.96$ where $\langle\mathit{\Delta}S\rangle_{\delta}/S_{0}$ is estimated to be larger than unity by using Eq.~\eqref{eq:exponentialscaling}. On the other hand, it is negligible for $Q\ll(2\pi)^2/\ln 2$: For instance, it becomes less than $10^{-10}$ for $Q<1$. This result means that the direct detectability of $S_0$ in noise measurements with limited resolution only depends on $Q$. 

The intrinsic distribution of the transferred particles through a resonant level continuously changes with the change in the parameters of system, e.g. Gaussian for $k_{B}T\to0$ in the equilibrium perfect transmission and bi-poissonian for $r\to0$ in the equilibrium with $\varepsilon_{0}=0$~\cite{Bagrets:2003,Levitov:2004}. The diversity in the distributions seems to be, however, irrelevant for the scaling feature of the deviation from the J-N relation. Our calculation indeed shows that the same exponential and the square root dependences represented by the universal coefficients and the exponent even when we change the parameters of the system, implying that the scaling behavior is universal not only in this specific distribution but also in other types of the distributions. 

To confirm our conjecture analytically, we use the following general cumulant generating function $\mathcal{C}_{\rm G}(\lambda)$,
\begin{align}
\mathcal{C}_{\rm G}(\lambda)\equiv \sum_{n=1}^{\infty}\frac{\kappa_{n}}{n!}(i\tilde{\lambda})^n
\label{eq:generalcumulant}
\end{align}
where $\tilde{\lambda}\equiv \lambda +2\pi\lfloor\lambda/2\pi+1/2\rfloor$ with $\lfloor\cdots\rfloor$ being the floor function. The periodicity of $\mathcal{C}_{\rm G}(\lambda)$ in $\lambda$ is crucial to ensure the integer value of the transferred electron number. We assume that the average and the variance of the distribution are given by $\kappa_{1}=I_{0}/e\varGamma$ and $\kappa_{2}=S_{0}/e^2\varGamma$, respectively. Substituting $\mathcal{C}_{\rm G}(\lambda)$ instead of $\mathcal{C}_{0}(\lambda)$ in Eq.\eqref{eq:averagednoise} for $\Delta>2$, we obtain the following equation,
\begin{align}
&\langle\mathit{\Delta}S\rangle_{\delta}\notag\\
&=-\frac{(e\Delta)^2\mathit{\Gamma}}{2\pi^2}\sum_{m\ge 1}\frac{\exp\big[{\cal T}\mathit{\Gamma}{\cal C}_{\rm G}^{{\rm sym}}(\frac{2\pi m}{\Delta})\big]{\cal C}_{\rm G}^{{\rm sym}}(\frac{2\pi m}{\Delta})}{m^2}\notag\\
&=-\frac{e^2\mathit{\Gamma}}{2\pi^2}\Big[\sum_{1\le n<\frac{\Delta}{2}}\frac{\pi^2}{\sin^2(\frac{\pi n}{\Delta})}\exp\big[{\cal T}\mathit{\Gamma}{\cal C}_{\rm G}^{{\rm sym}}(\frac{2\pi n}{\Delta})\big]{\cal C}_{\rm G}^{{\rm sym}}(\frac{2\pi n}{\Delta})\notag\\
&\qquad+\delta_{\Delta \bmod 2,0}\frac{\pi^2}{2}\exp\big[{\cal T}\mathit{\Gamma}{\cal C}_{\rm G}^{{\rm sym}}(\pi)\big]{\cal C}_{\rm G}^{{\rm sym}}(\pi)\Big]\notag\\
\label{eq:generalexcessnoise}
\end{align}
where ${\cal C}_{\rm G}^{{\rm sym}}(\lambda)\equiv{\cal C}_{\rm G}(\lambda)+{\cal C}_{\rm G}(-\lambda)$.

First we derive the exponential form emerging at $Q<1$. Using the expansion Eq.~\eqref{eq:generalcumulant}, Eq.~\eqref{eq:generalexcessnoise} is given by
\begin{align}
&\langle\mathit{\Delta}S\rangle_{\delta}\notag\\
&=2S_{0}\sum_{1\le n<\frac{\Delta}{2}}[1+{\cal O}((\frac{n}{\Delta})^2)]\exp\big[-\frac{(2\pi n)^2}{Q}[1+{\cal O}((\frac{n}{\Delta})^2)]\big]\notag\\
&\quad+\delta_{\Delta \bmod 2,0}\frac{S_{0}\pi^2}{4}\exp\big[-\frac{(\pi\Delta)^2}{Q}[1-\frac{2\pi^2\kappa_{4}}{4!\kappa_{2}}+\cdots]\big]\notag\\
&\quad\times[1-\frac{2\pi^2\kappa_{4}}{4!\kappa_{2}}+\cdots]\notag\\
&\sim 2S_{0}\exp\big[-(2\pi)^2/Q\big] \quad (Q\ll1).
\label{eq:estimation_exponential}
\end{align}
Hence, the deviation from the J-N relation is evaluated for $Q\ll1$ as Eq.~\eqref{eq:exponentialscaling} with $A=2$ and $\gamma=(2\pi)^2$ as is already mentioned.

Next we derive the square root dependence for $Q\gg1$. Since the square root dependence emerges only for $\Delta\gg1$, we evaluate the sum in Eq.~\eqref{eq:generalexcessnoise} with using the integral as
\begin{align}
&\langle\mathit{\Delta}S\rangle_{\delta}\notag\\
&\simeq-\frac{e^2\mathit{\Gamma}\Delta}{2\pi}\int_{0}^{\pi/2}\frac{dx}{\sin^{2}(x)}\exp\big[{\cal T}\mathit{\Gamma}{\cal C}_{\rm G}^{{\rm sym}}(2x)\big]{\cal C}_{\rm G}^{{\rm sym}}(2x)\notag\\
&= \frac{2e^2\mathit{\Gamma}\Delta\kappa_{2}}{\pi}\int_{0}^{\pi/2}dx\frac{x^2(1-\frac{2\kappa_{4}(2x)^2}{\kappa_{2}4!}+\cdots)}{\sin^{2}(x)}\notag\\
&\quad\times\exp\big[-{\cal T}\mathit{\Gamma}\kappa_{2}(2x)^2(1-\frac{2\kappa_{4}(2x)^2}{\kappa_{2}4!}+\cdots)\big]\notag\\
&\sim \frac{2e^2\mathit{\Gamma}\Delta\kappa_{2}}{\pi}\int_{0}^{\infty}dx\exp\big[-4{\cal T}\mathit{\Gamma}\kappa_{2}x^2\big] \quad(Q\ll2\pi^2\Delta^2)\notag\\
&= \sqrt{S_{\rm{}M}S_{0}/4\pi}.
\label{eq:estimation_squareroot}
\end{align}
For $1\ll Q\ll2\pi^2\Delta^2$, the deviation from the J-N relation, therefore, follows Eq.~\eqref{eq:squarerootscaling} with $B=1/\sqrt{4\pi}$. $A=2$, $B=1/\sqrt{4\pi}$, and $\gamma=(2\pi)^2$ perfectly agree with our numerical results (see Fig.~\ref{fig2}).

\begin{figure}[tb]
\begin{center}
\includegraphics[width=86mm]{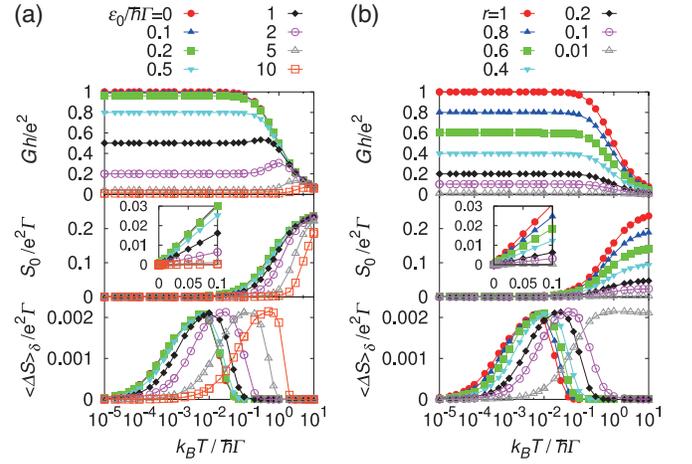}
\end{center}
\caption{
{(color online). Conductance $G$, intrinsic noise $S_{0}$, and excess noise $\langle\mathit{\Delta}S\rangle_{\delta}$ as a function of temperature $T$.} The measurement parameters are fixed at ${\cal{}T}\mathit{\Gamma}=1000$ and $\Delta=10$. $r=1$ and several choices of $\varepsilon_{0}$ are used in (a), and $\varepsilon_{0}=0$ and several choices of $r$ are used in (b). The insets show the enlarged plots of $S_{0}/{\rm e^2}\mathit{\Gamma}$.}
\label{fig4}
\end{figure}

This proof supports that the scaling functions represented by the exponential dependence Eq.~\eqref{eq:exponentialscaling} and the square root dependence Eq.~\eqref{eq:squarerootscaling} are universal irrespective of the details of the system. Therefore, this scaling should hold in general mesoscopic systems that are described by the Hamiltonian~\eqref{eq:Hamiltonian}, e.g. the quantum dot system in the Coulomb blockade regime~\cite{Bagrets:2003,Utsumi:2006} and in the presence of the energy dissipation by the electron-phonon coupling~\cite{Avriller:2009}. It also supports our expectation that the scaling does not directly depend on the internal system parameters specific to the present model.

In the following, we see the $\varepsilon_{0}$ and $r$-dependences of the conductance $G$, the intrinsic noise $S_{0}$, and the excess noise $\langle\mathit{\Delta S}\rangle_{\delta}$ in Fig.~\ref{fig4}. It is seen that all these transport quantities are strongly dependent on $\varepsilon_{0}$ and $r$. Since $G$ and $S_{0}$ are only determined by the system parameters, the characteristic temperature of those quantities is given by $\hbar\mathit{\Gamma}/k_{\rm{}B}$. For $k_{\rm{}B}T/\hbar\mathit{\Gamma}\ll 1$, $G$ takes a constant value and $S_{0}$ shows a simple linear dependence on $T$ expected from the J-N relation. While, $\langle\mathit{\Delta S}\rangle_{\delta}$ shows a strong temperature dependence even for $k_{\rm{}B}T/\hbar\mathit{\Gamma}\ll 1$ because it also depends on the measurement parameters, $\cal{}T$ and $\Delta$.

\begin{figure}[tb]
\begin{center}
\includegraphics[width=86mm]{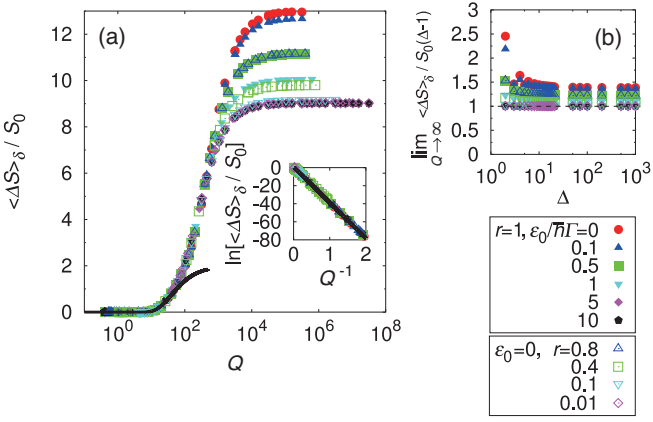}
\end{center}
\caption{(color online). (a) Ratio of excess and intrinsic noises at $V=0$, $\langle\mathit{\Delta}S\rangle_{\delta}/S_{0}$, as a function of $Q\equiv S_{\rm{}M}/S_{0}$ where $S_{\rm{}M}\equiv(e\Delta)^2/{\cal{}T}$ for several choices of ($\varepsilon_0,r$). ${\cal{}T}=1000$ and $\Delta=10$ is used for calculation. The black solid line indicates the universal exponential $A\exp[-\gamma/Q]$, with $A=2$ and $\gamma=(2\pi)^2$. The inset shows the linear dependence of the logarithm of the ratio on $Q^{-1}$. (b) Saturation value of the ratio of noises. The dashed line represents the lower bound of the saturation value.}
\label{fig5}
\end{figure}

Though it is seemingly difficult to find the universal relation between the transport properties for the different values of $\varepsilon_{0}$ and $r$, the scaling behavior is again confirmed even in the case. In Fig.~\ref{fig5}(a), it is also found the universality of the exponential form for $Q\ll10^2$. The saturation value of the deviation at high $Q$ stays at the order of $\Delta$ but weakly dependent on the system parameters. Figure~\ref{fig5}(b) shows the $\Delta$ dependence of the saturation value at sufficiently high $Q\gg{}4\pi\Delta^2$, where the lower bound of the saturation value is found,
\begin{equation}
\lim_{Q\to\infty}\langle\mathit{\Delta}S\rangle_{\delta}/S_{0}\ge\Delta-1.
\end{equation}
Hence, $S$ is always larger than $S_{0}\Delta$ in the limit of $Q\to\infty$.

\section{Comparison between theory and experiment}
In this section, we estimate realistic and presently accessible measurement parameters, ${\cal T}$ and $\Delta$, from an available measurement device.
In our two-point measurement scheme, the current is obtained by measuring the net transferred particle number within the measurement time, ${\cal T}$. 
Although the averaged current is precisely measurable for any choice of the parameters as discussed above, the rigorous value is obtained only when the average is given from the measurement performed infinitely many times.
When we consider the case of a single measurement, however, the measurement parameters should give a limit of available information about the current.

If the current is fluctuating with a frequency $f$, the detectability of the current must be crucially dependent on the measurement time $\cal T$. For $2f>{\cal T}^{-1}$, we hardly obtain the signal from the single measurement because the net transferred particle number within $\cal T$ is almost zero in our model. Therefore, we estimate the measurement time from the maximum detectable frequency in the actual single measurement, $f_{\rm max}$, as ${\cal T}=(2f_{\rm max})^{-1}$. In addition, the amplitude of the sinusoidal current with a frequency, $f_{\rm max}$, is important for the detectability. $\Delta$ specifies the detectable difference of the particle numbers at the initial and final states in the two-point measurement. If the net change of the number is less than $\Delta$, we have no meaningful signal in the single measurement. Hence, the minimum amplitude of the detectable sinusoidal current $I_{\rm min}$, with the frequency of $f_{\rm max}$ in the single measurement may give the estimation of $\Delta$ as $\Delta=\int_{0}^{\cal T}I_{\rm{}min}\sin(2\pi f_{\rm max}t)dt/e=I_{\rm min}/e\pi f_{\rm max}$.

\begin{figure}[tb]
\begin{center}
\includegraphics[width=86mm]{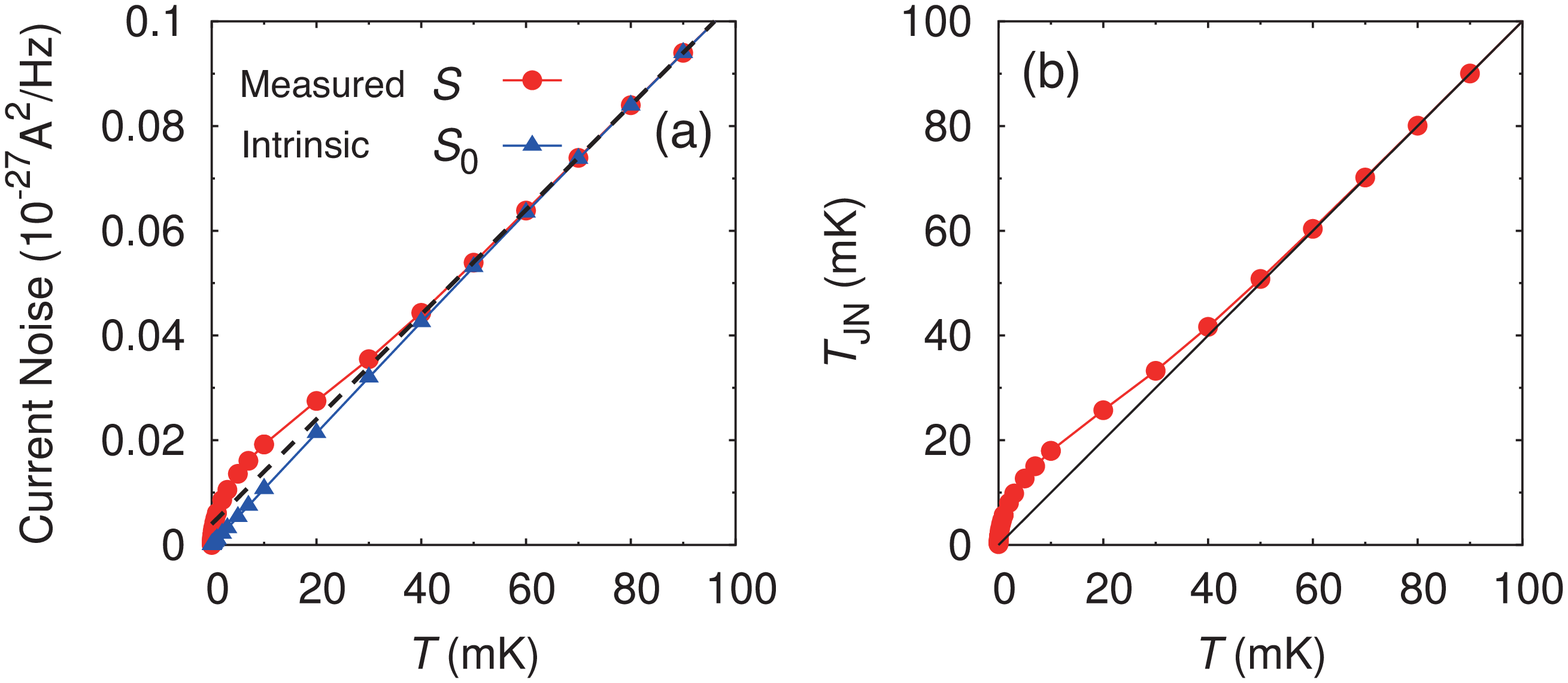}
\end{center}
\caption{(color online). (a) Current noise at $V=0$ as a function of temperature $T$. The parameters are $\hbar\mathit{\Gamma}/k_{\rm{}B}=1{\rm{}K}$, $\varepsilon_{0}=0$, $r=1$, ${\cal{}T}=1\mu{\rm{}s}$, and $\Delta=130$, which leads $S_{\rm{}M}=0.43$ $(10^{-27}{\rm{}A}^2{\rm{}Hz}^{-1})$. The dashed line indicates the fitted line for the measured noise from 50mK to 100mK, $aT+b$. $a=1.00$ ($10^{-27}{\rm{}A}^2{\rm{}Hz}^{-1}{\rm{}K}^{-1}$) is slightly smaller than the expected value for the intrinsic noise at low temperatures, $2k_{\rm{}B}e^2/h\simeq{}1.07$ ($10^{-27}{\rm{}A}^2{\rm{}Hz}^{-1}{\rm{}K}^{-1}$). $b=3.93$ (${}10^{-30}{\rm{}A}^2{\rm{}Hz}^{-1}$).
(b) Noise temperature $T_{\rm{}JN}$ plotted versus $T$. The parameters are the same as those in (a). The solid line shows $T_{\rm{}JN}=T$.}
\label{fig6}
\end{figure}

In the actual measurement of current through a mesoscopic device, the signal of current is enhanced via an amplifier because it is too weak to be directly measured with normal ammeters. Amplifiers have two significant parameters: The maximum detectable frequency, $f_{\rm amp}$, and the input current noise, $i_{n}$, which has the dimension of $\rm{}A/\sqrt{Hz}$. Since the precision of the current measurement is limited mainly by amplifiers, we connect our model parameters with those of an amplifier. Since the maximum frequency of the detectable current, $f_{\rm max}$, is supposed to be given by $f_{\rm amp}$, the measurement time is estimated as
\begin{equation}
{\cal T}=(2f_{\rm amp})^{-1}.
\label{eq:estimationT}
\end{equation}
The input current noise limits the amplitude of the detectable current. To obtain meaningful information in a single measurement, the input sinusoidal current with a frequency of $f_{\rm amp}$ must have the amplitude larger than $i_{n}\sqrt{f_{\rm amp}}$, which leads to $I_{\rm min}=i_{n}\sqrt{f_{\rm amp}}$. Hence, we estimate $\Delta$ as 
\begin{equation}
\Delta=i_{n}/e\pi\sqrt{f_{\rm amp}}.
\label{eq:estimationD}
\end{equation}
More concretely, we estimate ${\cal T}$ and $\Delta$ from the amplifier of CA-554F2 manufactured by NF Corporation in Japan. CA-554F2 is one of the best amplifiers on the market, which has $f_{\rm amp}=500{\rm KHz}$ and $i_{n}=45{\rm fA/\sqrt{Hz}}$. Substituting these parameters into Eq. (\ref{eq:estimationT}) and Eq. (\ref{eq:estimationD}), we obtain ${\cal T}\simeq 1\mu{\rm s}$ and $\Delta\simeq 130$.~\cite{comments}

In Fig.~\ref{fig6}(a), the measured and intrinsic noises are plotted versus the temperature for realistic model parameters, $\hbar\mathit{\Gamma}/k_{\rm{}B}=1{\rm{}K}$, $\varepsilon_{0}=0$, and $r=1$. Note that for $T\ll1{\rm{}K}$, the model effectively describes the single channel QPC with perfect transmission. We use ${\cal{}T}=1\mu{\rm{}s}$ and $\Delta=130$. At temperatures higher than 50mK, $S$ shows a clear linear dependence on temperature and takes nearly the same value as $S_0$. While, $S$ deviates form $S_0$ and makes a hump at lower temperatures below 50mK. These features are qualitatively consistent with the experiment~\cite{Hashisaka:2008ef}.

Finally, we show the noise temperature in the realistic conditions. Because the noise temperature, $T_{\rm{}JN}$, is explicitly written as 
\begin{equation}
T_{\rm{}JN}\equiv{}S/2k_{\rm{}B}G=T(1+\langle\mathit{\Delta}S\rangle_{\delta}/S_{0}),
\end{equation}
it is always larger than the thermodynamic temperature, $T$. Figure~\ref{fig6}(b) shows $T_{\rm{}JN}$ as a function of $T$ for the same parameters as those in Fig.~\ref{fig6}(a). The disagreement of $T_{\rm{}JN}$ with $T$ appears below 50mK, which is also consistent with the experiments~\cite{Hashisaka:2008ef,Nakamura:2010hn,Webb:1973ej}. This result indicates that the intrinsic temperature may not be obtained in the Johnson noise thermometry at very low temperatures even if $T_{\rm{}JN}\simeq T_{\rm{}ref}$ holds at higher temperatures. 

\begin{figure}[tb]
\begin{center}
\includegraphics[width=80mm]{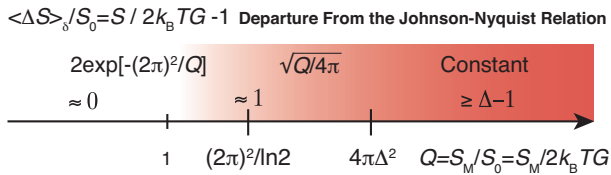}
\end{center}
\caption{(color online). Schematic illustration of departure from Johnson-Nyquist relation for a large $\Delta\gg1$. The universal departure starts at the essential singular point of the exponential function, $2\exp[-(2\pi)^2/Q]$, which followed by the square root dependence $\sqrt{Q/4\pi}$.}
\label{fig7}
\end{figure}

\section{Summary and Prospect}
We summarize our findings as schematic in Fig.~\ref{fig7}, where the universal departure from the J-N relation is characterized by the single parameter $Q$. Moreover, the departure starts with a universal function characterized by an exponential form, $2 \exp[-(2\pi)^2/Q]$, when the ideal resolution becomes lost from $Q=0$ where the function has the essential singularity. Then, it is followed by the square root growth, $\sqrt{Q/4\pi}$. In addition to the present proposal, there exist other possible scenarios to explain the experimental anomalous noise enhancement including the heat leak. The smoking gun to prove our proposal is whether the scaling behavior is satisfied or not. It is desired to test in experiments. In this paper, we have focused on the J-N relation within the linear response. Even for the nonlinear regime, similar puzzles of the deviation from the fluctuation theorem~\cite{Nakamura:2010hn} and the discrepancy of the shot noise between theory and experiment are known~\cite{Yamauchi:2011cq}. The resolution effects may also give us a clue to resolve them. More generally, our results may propose the necessity of amending naive accounts of resolution effects in widespread instruments based on the fluctuation-dissipation theorem such as nuclear magnetic resonance, X-ray scattering, neutron scattering, and photoemission. 

\section*{ACKNOWLEDGEMENT}
The authors are grateful to T.~Fujisawa, M.~Hashisaka, K.~Inaba, T.~Kato, K.~Kobayashi, S.~Morita, A.~Oguri, K.~Saito, R.~Sakano, Y.~Utsumi, Y.~Watanabe, and M.~Yamashita for fruitful discussions. This work is financially supported by Grant-in-Aid for Scientific Research (No. 22104010, and 22340090) from MEXT, Japan. This work is financially supported by MEXT HPCI Strategic Programs for Innovative Research (SPIRE) and Computational Materials Science Initiative (CMSI).


\begin{thebibliography}{100}
\bibitem{Levitov:1993ma} L.~Levitov and G.~Lesovik, JETP Lett. \textbf{58}, 230 (1993).
\bibitem{Levitov:1996ie} L.~Levitov, H.~Lee, and G.~Lesovik, J. Math. Phys. \textbf{37}, 4845 (1996).
\bibitem{Nazarov:2003} \textit{Quantum Noise in Mesoscopic Physics, Vol. 97 of NATO Science Series II: Mathematics, Physics and Chemistry}, edited by Yu.~V.~Nazarov (Kluwer, Dordrecht, 2003).
\bibitem{Esposito:2009zz} M.~Esposito, U.~Harbola, and S.~Mukamel, Rev. Mod. Phys. \textbf{81}, 1665 (2009).
\bibitem{Naaman:2006} O.~Naaman and J.~Aumentado, Phys. Rev. Lett. \textbf{96}, 100201 (2006).
\bibitem{Utsumi:2010} Y.~Utsumi, D.~S.~Golubev, M.~Marthaler, T.~Fujisawa, and G.~Sch\"{o}n, in \textit{Perspectives of Mesoscopic Physics--Dedicated to Yoseph Imry's 70th Birthday}, edited by A.~Aharoni and O.~Entin-Wohlman (World Scientific, Singapore, 2010), pp. 397-414.
\bibitem{Bednorz:2008} A.~Bednorz and W. Belzig, Phys. Rev. Lett. 101, 206803 (2008).
\bibitem{Johnson:1927tu} J.~B.~Johnson, Nature \textbf{119}, 50 (1927); Phys. Rev. \textbf{32}, 97 (1928).
\bibitem{Nyquist:1928wx} H.~Nyquist, Phys. Rev. \textbf{32}, 110 (1928).
\bibitem{Callen:1951wg} H.~B.~Callen and T.~A.~Welton, Phys. Rev. \textbf{83}, 34 (1951).
\bibitem{Kubo:1957wk} R.~Kubo, J. Phys. Soc. Jpn. \textbf{12}, 570 (1957).
\bibitem{White:1996wr} D.~R.~White, R.~Galleano, A.~Actis, H.~Brixy, M.~De~Groot, J.~Dubbeldam, A.~L.~Reesink, F.~Edler, H.~Sakurai, and R.~L.~Shepard, Metrologia \textbf{33}, 325 (1996).
\bibitem{Reznikov:1995us} M.~Reznikov, M.~Heiblum, H.~Shtrikman, and D.~Mahalu, Phys. Rev. Lett. \textbf{75}, 3340 (1995).
\bibitem{depicciotto:1997dk} R.~de~Picciotto, M.~Reznikov, M.~Heiblum, V.~Umansky, G.~Bunin, and D.~Mahalu, Nature \textbf{389}, 162 (1997).
\bibitem{Saminadayar:1997tl} L.~Saminadayar, D.~C.~Glattli, Y.~Jin, and B.~Etienne, Phys. Rev. Lett. \textbf{79}, 2526 (1997).
\bibitem{Lefloch:2003fp} F.~Lefloch, C.~Hoffmann, M.~Sanquer, and D.~Quirion, Phys. Rev. Lett. \textbf{90}, 067002 (2003).
\bibitem{Sela:2006kq} E.~Sela, Y.~Oreg, F.~von Oppen, and J.~Koch, Phys. Rev. Lett. \textbf{97}, 086601 (2006).
\bibitem{Zarchin:2008gq} O.~Zarchin, M.~Zaffalon, M.~Heiblum, D.~Mahalu, and V.~Umansky, Phys. Rev. B \textbf{77}, 241303(R) (2008).
\bibitem{Hashisaka:2008ef} M.~Hashisaka, Y.~Yamauchi, S.~Nakamura, S.~Kasai, K.~Kobayashi, and T.~Ono, J. Phys.: Conf. Ser. \textbf{109}, 012013 (2008).
\bibitem{Delattre:2009} T.~Delattre, C.~Feuillet-Palma, L.~G.~Herrmann, P.~Morfin, J.~M.~Berroir, G.~F\`{e}ve, B.~Plaais, D.~C.~Glattli, M.~S.~Choi, C.~Mora, and T.~Kontos, Nature Physics \textbf{5}, 208 (2009).
\bibitem{Yamauchi:2011cq} Y.~Yamauchi, K.~Sekiguchi, K.~Chida, T.~Arakawa, S.~Nakamura, K.~Kobayashi, T.~Ono, T.~Fujii,and R.~Sakano, Phys. Rev. Lett. \textbf{106}, 176601 (2011).
\bibitem{Tobiska:2005ht} J.~Tobiska and Y.V.~Nazarov, Phys. Rev. B \textbf{72}, 235328 (2005).
\bibitem{Saito:2008hs} K.~Saito and Y.~Utsumi, Phys. Rev. B \textbf{78}, 115429 (2008).
\bibitem{Nakamura:2010hn} S.~Nakamura, Y.~Yamauchi, M.~Hashisaka, K.~Chida, K.~Kobayashi, T.~Ono, R.~Leturcq, K.~Ensslin, K.~Saito, Y.~Utsumi, et~al., Phys. Rev. Lett. \textbf{104}, 080602 (2010).
\bibitem{Webb:1973ej} R.~A.~Webb, R.~P.~Giffard, and J.~C.~Wheatley, J. Low Temp. Phys. \textbf{13}, 383 (1973).
\bibitem{Davies:1970ux} E.~B.~Davies and J.~T. Lewis, Commun. Math. Phys. \textbf{17}, 239 (1970).
\bibitem{Kraus:1971wd} K.~Kraus, Ann. Phys. (N.Y.) \textbf{64}, 311 (1971).
\bibitem{Fujisawa:2006jf} T.~Fujisawa, T.~Hayashi, R.~Tomita, and Y.~Hirayama, Science \textbf{312}, 1634 (2006).
\bibitem{Gustavsson:2006jm} S.~Gustavsson, R.~Leturcq, B.~Simovi\v{c}, R.~Schleser, T.~Ihn, P.~Studerus, K.~Ensslin, D.~C.~Driscoll, and A.~C.~Gossard, Phys. Rev. Lett. \textbf{96}, 076605 (2006).
\bibitem{Gustavsson:2007} S.~Gustavsson, R.~Leturcq, T.~Ihn, K.~Ensslin, M.~Reinwald, and W.~Wegscheider, Phys. Rev. B \textbf{75}, 075314 (2007).
\bibitem{Kung:2012ct} B.~K{\"u}ng, C.~R{\"o}ssler, M.~Beck, M.~Marthaler, D.S.~Golubev, Y.~Utsumi, T.~Ihn, and K.~Ensslin, Phys. Rev. X, \textbf{2}, 011001 (2012).
\bibitem{DiCarlo:2006} L.~DiCarlo, Y.~Zhang, D.~T.~McClure, C.~M.~Marcus, L.~N.~Pfeiffer, and K.~W.~West, Rev. Sci. Instrum. \textbf{77}, 073906 (2006).
\bibitem{Kindermann:2003th} M.~Kindermann and Y.~V.~Nazarov, in \emph{Quantum Noise in Mesoscopic Physics}, edited by Y.~V.~Nazarov (Kluwer, Dordrecht, 2003), pp. 403--427.
\bibitem{Kamenev:2005vu} A.~Kamenev, in \emph{Nanophysics: Coherence and Transport}, edited by H.~Bouchiat, Y.~Gefen, S.~Gu{\'e}ron, G.~Montambaux, and J.~Dalibard (Elsevier, Amsterdam, 2005), pp. 177--246.
\bibitem{Bagrets:2003} D.~A.~Bagrets and Y.~V.~Nazarov, Phys. Rev. B \textbf{67}, 085316 (2003).
\bibitem{Utsumi:2006} Y.~Utsumi, D.~S.~Golubev, and G.~Sch\"{o}n, Phys. Rev. Lett. \textbf{96}, 086803 (2006).
\bibitem{Avriller:2009} R.~Avriller and A.~Levy~Yeyati, Phys. Rev. B \textbf{80}, 041309 (2009).
\bibitem{Levitov:2004} L.~S.~Levitov and M.~Reznikov, Phys. Rev. B \textbf{70}, 115305 (2004).
\bibitem{comments} This should, however, be regarded as a rough estimate. To investigate the precise correspondence of the measurement parameters between theory and experiment, we further need more quantitative estimate of the resolution together with the data on the deviation of the J-N relation and its extensions in the nonlinear response.
\end{thebibliography}
\end{document}